\RequirePackage{fix-cm}
\documentclass{svjour3}                     % onecolumn (standard format)
\smartqed  % flush right qed marks, e.g. at end of proof
\usepackage{graphicx}
\usepackage{latexsym}
\usepackage{amsmath}
\usepackage{amssymb}
\usepackage{amsfonts}
\usepackage{color}
\usepackage{bm}% bold math
\usepackage{verbatim}
\begin{document}

\title{Quasiclassical theory of spin dynamics in superfluid $^3$He: kinetic equations in the bulk  
and spin response of surface Majorana states}



\author{M.A.~Silaev}


\institute{M.A.~Silaev \at
              Department of
 Physics and Nanoscience Center, University of Jyv\"askyl\"a, \\
 P.O.
 Box 35 (YFL), FI-40014 University of Jyv\"askyl\"a, Finland \\
              %  
              %first address \\
              \email{mikesilaev@gmail.com} 
}

\date{Received: date / Accepted: date}

\maketitle

\begin{abstract}
 We develop a theory based on the formalism of quasiclassical Green's functions to study 
 the spin dynamics in superfluid $^3$He. First, we derive 
 kinetic equations for the spin-dependent distribution function in the bulk superfluid
 reproducing the results obtained earlier without quasiclassical approximation. 
 Then we consider spin dynamics near the surface of fully gapped $^3$He-B phase taking into account spin relaxation 
 due to the transitions in the spectrum of localized fermionic states. 
 The lifetimes of longitudinal and transverse spin waves are calculated taking 
 into account the 
  Fermi-liquid corrections which lead to a crucial modification of 
  fermionic spectrum and spin responses.
\keywords{Superfluid $^3$He \and Magnetic resonance \and Majorana states}
\end{abstract}

\section{Introduction}
 The theory of spin dynamics in superfluid $^3$He has been developed for several decades since the pioneering
 works of Leggett \cite{Leggett1972,Leggett1973,Leggett1975a}, where the phenomenological equations were 
 formulated explaining shifts of the transverse nuclear magnetic resonance mode
  and predicting longitudinal resonance in the B phase \cite{Osheroff1972,Osheroff1972a,Osheroff1974}. 
  To study spin relaxation, that is the width 
 of the NMR signal, Leggett and Takagi \cite{Leggett1975,Leggett1977} introduced the two-fluid model
 which yields qualitatively the same results as the kinetic theory 
 \cite{Combescot1974a,Combescot1974,Combescot1975,Combescot1976,Woelfle1976}.
 
  Nowadays, the most common approach to study non-equilibrium 
 states in different condensed matter systems including superconductors and Fermi superfluids is
 based on the quasiclassical Keldysh formalism. 
 In this way kinetic equations for spin-singlet superconductors were derived 
 \cite{Schmid1975,Larkin1969} and applied to study various problems 
 (see e.g. the book \cite{KopninBook} for review).
 Recently this theory has been extended 
 to study non-equilibrium spin states in spin-singlet superconducting materials 
 \cite{Bobkova2015,Silaev2015,SilaevRMP2017}.
 
 Interestingly enough, although the general quasiclassical approach  
 to superfluid $^3$He has been described \cite{Serene1983},
 kinetic equations for spin dynamics in spin-triplet superconductors/superfluids have never been derived using this technique.
 An attempt to apply quasiclassical methods to spin dynamics has been done in Ref.\cite{Eckern1981}.
 However, this work does not reproduce kinetic equations obtained without the quasiclassical approximation
 \cite{Combescot1974a,Combescot1974,Combescot1975,Combescot1976,Woelfle1976}, 
 essentially because several important terms have been omitted during the derivation.
 The purpose of the present paper is partially to close this gap.   
  
 Being powerful tools to study spin dynamics in the bulk,
 previous kinetic theories are not capable to describe the spin response of  
 localized quasiparticle surface states dubbed Andreev-Majorana fermions \cite{Volovik2010,Silaev2011a,SilaevVolovik2014,Mizushima2015}. 
 Recently the frequency-dependent linear spin response of Andreev-Majorana
  states was found \cite{Silaev2011a}.
 As noted in Ref.\cite{Taylor2015} to obtain the 
 total spin susceptibility it is necessary to take into account the 
 self-consistent response of the spin-triplet order parameter. 
  In the present paper we apply quasiclassical formalism to this problem, calculating the 
 coupled dynamics of spin waves and Andreev-Majorana surface states in the film of fully gapped superfluid $^3$He-B.
      This approach allows treating both the longitudinal and transverse magnetic resonances and 
      taking into account Fermi-liquid corrections
 which are quite important in $^3$He \cite{Abrikosov1958,Mizushima2012,Silaev2015}. 
 We show that these corrections
 can drastically change the spin response properties shifting the threshold absorption frequency of the surface states to values several 
 times larger than the basic Larmor frequency of magnetic precession. 
   
 \section{Keldysh formalism for non-equilibrium spin states}
 \label{Sec:KeldyshFormalism}
 
 \subsection{General equations}
 In general the spin density ${\bm S}$ can be written in terms of the Keldysh quasiclassical 
 Green's function (GF) as
 \begin{equation} \label{Eq:Spin}
 {\bm S}({\bm r},t) = 
 \frac{\hbar\nu_0}{16}  \int \frac{d\Omega_p}{4\pi} {\rm Tr}
 \left[ \hat\tau_3 \hat{\bm \sigma} \widehat g^K (t,t) \right]
 + {\bm S}^{(n)}_{eq} ,
 \end{equation}
 where $\nu_0$ is the normal-state density of states, $\tau_i$, $\sigma_i$ are Pauli matrices in Nambu and spin spaces,
 $\widehat g^K$ is the (2$\times$2 matrix) Keldysh component of the matrix quasiclassical Green's function 
 \begin{equation}
 \check{g} = \left(%
 \begin{array}{cc}
  \widehat{g}^R &  \widehat{g}^K \\
  0 &  \widehat{g}^A \\
 \end{array}\label{eq:GF0}
 \right)\; ,
 \end{equation}
 and $\widehat{g}^{R(A)}$ are the retarded (advanced) propagators.
 The addition term in (\ref{Eq:Spin}) compensates off-shell contributions 
 ${\bm S}^{(n)}_{eq} = \chi_n \bm {H}_{ext}$, where $\chi_n$ is the normal-state susceptibility and 
 $\bm {H}_{ext}$ is an external field. 

 In clean superfluid the matrix $\check g$ obeys the Keldysh-Eilenberger equation \cite{Eilenberger1968}
 \begin{equation}\label{Eq:Eilenberger}
 \bm {v_F\cdot\nabla} \check{g} +  \{\hat\tau_3\partial_t, \check{g} \}_t
 + i [\check \Sigma, \check{g} ]_t + i [\check V_Z, \check{g} ]_t = St (\check{g}),
 \end{equation}
 where $\bm v_F$ is the Fermi velocity and the commutator/anti-commutator operators are defined as
  \begin{align}
  & [X, g]_t= X(t_1) g(t_1,t_2)- g(t_1,t_2)X(t_2) \\
  & \{X, g\}_t= X(t_1) g(t_1,t_2)+ g(t_1,t_2) X(t_2) .
  \end{align}
  The l.h.s of Eq.(\ref{Eq:Eilenberger}) contains the spin-dependent Zeeman energy 
  $\check{V}_Z = \hat I_K \widehat{V}_Z $  where $\hat I_K$ is the unit 
  matrix in Keldysh space,  
  $\widehat{V}_Z = -\gamma \hat\tau_3(\hat{\bm \sigma}\cdot\bm{H}_{ext})/2 $, 
  where $\gamma$ is the gyromagnetic ratio
  of $^3$He nuclei. % and $\bm H_{ext}$ is the external magnetic field. 
  Similarly, the Keldysh component of the Hartree-Fock self-energy is absent and 
   the spectral components are given by the superposition of three terms:
  $\widehat \Sigma^{R(A)} = \widehat{\Sigma}_{FL} + \widehat{\Delta} + \widehat{\Sigma}_D $.
  %\end{equation}
  The first one is the Fermi-liquid self-energy, which we take in the form 
  describing the correction to the Zeeman field 
  $\widehat{\Sigma}_{FL} = \hat\tau_3 \gamma^2 (Z_0/8) (\hat{\bm \sigma}\cdot\bm{S})  /\chi_{n0} $,
  where $Z_0\approx -3$ is the Landau parameter, describing the enhancement of spin susceptibility, 
  $\chi_{n0} = \gamma^2\hbar^2 \nu_0/4$ is the normal-state susceptibility without corrections.
  As a result the effective magnetic field modified by Fermi-liquid corrections is given by 
   \cite{Leggett1977}
  \begin{equation} \label{Eq:Heff}
  {\bm H}_{eff}= {\bm H}_{ext} - \gamma (Z_0/4) {\bm S}/\chi_{n0} .
 \end{equation}    
  The self-energy $\widehat{\Sigma}$  contains off-diagonal gap and dipolar interaction operators.
  The gap is parametrized in terms of the spin vector $\bm{d}$ and is given by
  $\widehat{\Delta} = i \Delta_0 \hat\tau_1 (\hat{\bm \sigma}\cdot \bm{d}) $. 
  The general form of the dipolar interaction is not important for the present paper.   
  Similarly, we do not specify the particle-particle collision integral on the r.h.s. of Eq. (\ref{Eq:Eilenberger}).
  
  Equation \eqref{Eq:Eilenberger} is complemented by the normalization condition
  $\check g \circ\check g =1$ that allows writing the Keldysh component 
  \begin{equation}\label{Eq:parametrization}
  \widehat{g}^K (t_1,t_2)= \widehat{g}^R \circ \widehat{f} - \widehat{f} \circ \widehat{g}^A ,
  \end{equation}
  where $\widehat{f}=\widehat{f}(t_1,t_2)$ is the generalized distribution function and 
  the convolution product is defined as
  $ (A\circ B) (t_1,t_2) = \int dt A(t_1,t) B(t,t_2) $.
   
 %%%%%%%%%%%%%%%%%%%%%%%%%%%%%%%%%%%%%%%%%%%%%%%%%%%%%%%%%%          
 \subsection{Spin conservation} 
 The spin conservation law can be obtained from the general Keldysh-Eilenberger equation (\ref{Eq:Eilenberger})
 using the definition of spin density (\ref{Eq:Spin}) and the self-consistency relation
 for the order parameter which yields ${\rm Tr} \bm{\hat \sigma} \left[\widehat{\Delta}, \widehat{g}^K \right]=0$ .
 Moreover, the Fermi-liquid self-energy contribution drops out as well 
 ${\rm Tr} \bm{\hat \sigma} \left[\widehat{\Sigma}_{FL}, \widehat{g}^K \right]=0$. 
 Therefore multiplying the Keldysh component of Eq. (\ref{Eq:Eilenberger}) by 
 $\hat{\bm \sigma}$ and taking the trace we obtain the exact equation:
 \begin{equation}\label{Eq:SpinConservation}
 \nabla_k \bm{J}_{k} + \bm{\dot{S}} = \gamma \bm{S}\times \bm{H}_{ext} + \bm{R}_D,
 \end{equation}   
 where the components of the spin current $\bm{J}_{k}$ and the dipole 
 torque $\bm{R}_D$ are defined as follows 
  \begin{align} \label{Eq:SpinCurrent}
  & \bm {J}_{k} = \frac{\hbar\nu_0}{16} \int \frac{d\Omega_p}{4\pi} v_{F k} {\rm Tr}( \hat{\bm \sigma} \widehat g^K),
  \\ \label{Eq:RDgeneral}
  & \bm {R}_D = \frac{i\hbar\nu_0}{8} \int \frac{d\Omega_p}{4\pi} 
  {\rm Tr} ({\bm \sigma} [\widehat\Sigma_D, \widehat{g}^K]) .
  \end{align} 
  
 %%%%%%%%%%%%%%%%%%%%%%%%%%%%%%%%%%%%%%%%%%%%%%%%%%%%%%%%%%%%%%%%%%%%   
 \subsection{Rotating frame: general case}
  There are two possible types of driving terms in the kinetic equation (\ref{Eq:Eilenberger})
  which generate non-equilibrium states.
  One of them is the time-dependent external field 
  $\bm{H}_{ext}= \bm{H}_{ext} (t)$ generated by external sources. 
   The other driving term is given by the time-dependent rotating order parameter vector
  $\bm {d}(t) = \hat R^{-1} (t) \bm {d}_0(\bm k)$, where the rotation matrix $\hat R$ 
  is determined by its angle $\theta (t)$ and axis ${\bm n} (t)$:
  \begin{equation}\label{Eq:RotationMatrix}
  R_{ik} = \delta_{ik} + (n_in_k - \delta_{ik}) (1-\cos\theta)- \varepsilon_{ikl} n_l \sin\theta .
  \end{equation}
  In general the rotation matrix can be split $\hat R(t) = \hat R_{ac}(t) \hat R_0$ into the time-independent part $\hat R_0$,
  describing the static  order parameter configuration and the part $\hat R_{ac}(t)$ corresponding to the time-dependent 
   rotation (\ref{Eq:Rotation}).     
  To simplify the kinetic equation Eq.(\ref{Eq:Eilenberger}) 
  it is convenient to introduce the rotating frame by removing the time dependence of the order
  parameter with the help of the following SU(2) transformation
  \begin{align} \label{Eq:Rotation}
  \check{\tilde{g}} (t_1,t_2) = \hat U^\dagger (t_1) \check{g}(t_1,t_2) \hat U (t_2),
  %, \\
  %\widehat{\tilde{\Lambda}} (t) = \hat U^\dagger (t) \widehat{\Lambda} (t) \hat U (t)
  \end{align}
  where $\hat U (t)= e^{i\bm {\hat\sigma }\cdot\bm{\theta}(t)/2}$.
  This transformation rotates spin vectors
  $\hat U^\dagger \bm{\hat \sigma} \hat U=\hat R^{-1}_{ac} \bm{\hat \sigma}$
  so that the gap function becomes time-independent 
  $\hat{\tilde{\Delta}} = i\tau_1({\bm {\hat \sigma}}\cdot\bm {d}_0)$
  and the Zeeman energy in the rotating frame acquires a new term as follows
  \begin{equation}\label{Eq:SelfEnergyRotatingFrame}
  \widehat{V}_Z = - \frac{\gamma}{2}\hat\tau_3 ( \bm {\hat \sigma} \cdot \hat R_{ac} {\bm H}_{eff} )
  - i \hat\tau_3 \hat U^\dagger \partial_t\hat{U},
  \end{equation}
  where the effective field in the first term is given by Eq. (\ref{Eq:Heff}). Here the spin-dependent
  Fermi-liquid corrections are incorporated
  into the Zeeman term. 
 For small angles $\theta$ one can expand the rotation matrix 
 $\hat R_{ac} {\bm H}_{eff} \approx {\bm H}_{eff}  - {\bm H}_{eff}\times {\bm \theta}$,     
 and put $ \hat U^\dagger \partial_t\hat{U} \approx i {\bm {\hat \sigma}\cdot\partial_t\bm{\theta} }/2 $
 which allows rewriting the kinetic equation (\ref{Eq:Eilenberger}) separating the  
 time-independent term $\check\Lambda_0 =  \hat I_K \widehat \Lambda_0 $ and 
 the driving term $\check V_{Z1} = \hat I_K \widehat V_{Z1}$  as follows
 \begin{equation}\label{Eq:EilenbergerOrder1}
 i\bm {v_F\cdot\nabla} \check{\tilde{g}} +  i\{\hat\tau_3\partial_t, \check{\tilde{g}} \}_t
 - [\check \Lambda_0, \check{\tilde{g}} ]= [\check V_{Z1}, \check{\tilde{g}} ]_t +  St (\check{\tilde{g}} ) ,
 \end{equation}
 where 
 $\widehat \Lambda_0 = i\Delta_0 \hat\tau_1 ({\bm {\hat \sigma}}\cdot \bm d_0) - 
 \gamma \hat\tau_3 ({\bm {\hat \sigma}}\cdot\bm{H}^{(0)}_{eff}) /2$
 and $\check V_{Z1}(t) = - \gamma \hat\tau_3 ({\bm {\hat \sigma}}\cdot \bm{h})/2$.  
 The driving field is given by
 \begin{equation} \label{Eq:TimeDependentField}
 \bm{h} (t)= \bm{\tilde H}_{eff} + {\bm \theta}\times{\bm H}^{(0)}_{eff} - 
 \partial_t \bm {\theta}/\gamma,
 \end{equation}
 where $\bm{H}^{(0)}_{eff}$ and $\bm{\tilde H}_{eff}$ are the constant and time-dependent 
 parts of the effective field (\ref{Eq:Heff}) 
 in the rotating frame.
 Below we will use Eq. (\ref{Eq:EilenbergerOrder1}) to find the first-order 
 non-equilibrium corrections to the
 Keldysh function in various situations. Namely at first we will derive kinetic equations
 for the spin distribution function describing  the bulk NMR in superluid $^3$He. 
 Second, we will calculate the spin response in the presence of Majorana surface states 
 in fully gapped B phase. 
         
  %%%%%%%%%%%%%%%%%%%%%%%%%%%%%%%%%%%%%%%%%%%%%%555
 
   \section{ Spin dynamics in the bulk superfluid $^3$He}
  In this section we apply the general formalism described above to derive kinetic equations
  for the spin-dependent distribution function which were obtained before without quasiclassical 
  approximation \cite{Combescot1974a,Combescot1974,Combescot1975,Combescot1976,Woelfle1976}. 
  The advantage of the quasiclassical approach is that the derivation becomes much simpler and one can clearly 
  understand the approximations made.
       
 %%%%%%%%%%%%%%%%%%%%%%%%%%%%%%%%%%%%%%%%%
 \subsection{Mixed representation}
   
  We will use the mixed representation of Green's functions 
  $ \check g(t_1,t_2)= e^{i\varepsilon (t_2-t_1)} \check g(\varepsilon,t)$ where $t=(t_1+t_2)/2$.
  Then with the help of the gradient expansion 
  $ \left[\check X, \check g\right]_t= [ \check X, \check g] - i \{ \partial_t \check X, \partial_\varepsilon \check g \}/2  $
  the Keldysh component of Eq. (\ref{Eq:Eilenberger}) can be written in the form
 \begin{equation}\label{Eq:EilenbergerExp}
 \bm {v}_F\cdot{\bm\nabla} \widehat{g}^K - i\varepsilon [\hat\tau_3, \widehat g^K]  +  \frac{1}{2}\{\hat\tau_3, \partial_t \widehat g^K \}
 + i [ \widehat\Lambda, \widehat g^K] + \frac{1}{2}\{ \partial_t \widehat\Lambda, \partial_\varepsilon \widehat g^K \}= St (\check{g}),
 \end{equation}
 where we denote $\widehat{\Lambda} = \widehat{\Sigma} + \widehat V_Z$. 
 To describe non-equilibrium spin states we use the 
 parametrization (\ref{Eq:parametrization}) of the Keldysh function. 
 Using the gradient expansion in the mixed representation the relation
  (\ref{Eq:parametrization}) can be written as follows
   \begin{equation} \label{Eq:gKgradientExp}
  \widehat{g}^K=
  \widehat{g}^R  \hat{f} - \hat{f} \widehat{g}^A
  -\frac{i}{2}\left(  \partial_t\widehat{g}^R  \partial_\varepsilon\hat{f} + 
  \partial_\varepsilon\hat{f} \partial_t \widehat{g}^A \right) +
  \frac{i}{2}\left( \partial_\varepsilon\widehat{g}^R \partial_t\hat{f}_1 + \partial_t\hat{f}_1
  \partial_\varepsilon \widehat{g}^A \right).
  \end{equation}
  We represent the distribution function in the form
 $\hat f (\varepsilon,t) = f_0(\varepsilon) + \hat f_1 (\varepsilon, t)$,
 where $f_0=\tanh(\varepsilon/2T)$ is the equilibrium part and 
 $\hat f_1= ({ \bm{\hat \sigma}\cdot \bm {f_T}})$ describes the spin non-equilibrium.
 The last two terms in Eq.(\ref{Eq:gKgradientExp}) containing the time  derivative
 $\partial_t  \hat f_1$ can be neglected
 provided that the frequency $\omega$ defined by the driving term is much smaller as compared to 
 the typical energy scale, which is of the order of the energy gap $\Delta$. 
 In this case $\hat f_1 \propto \omega$ and $\partial_t  \hat f_1 \propto \omega^2$, so that the last two terms in 
 Eq.(\ref{Eq:gKgradientExp}) are of the higher-order in the small parameter $\omega/\Delta$.
 % 
 %Thus we obtain $  \widehat{g}^K= \widehat{g}^{RA} f_0 + (\widehat{g}^R  \hat{f}_1 - \hat{f}_1 \widehat{g}^A )  $. 
 When we substitute the expansion 
 $ \widehat{g}^K = \widehat{g}^R  \hat{f}- \hat{f} \widehat{g}^A  $
 to (\ref{Eq:EilenbergerExp}) some terms cancel due to the Eilenberger equation (\ref{Eq:Eilenberger})
 for spectral components $\widehat{g}^{R,A}$.
  Thus we are left with a kinetic equation for the spin-dependent distribution function
  %%%%%%%%%%%%%%%%%%%%%%%%%%%%%%%%%%%%%%%%%%
     \begin{align}\label{Eq:GeneralKinetic}
     \bm {v}_F\cdot \bm{\nabla }(\widehat g^R \hat f_1 - \hat f_1 \widehat g^A )+
     \frac{1}{2}\{\hat\tau_3, (\widehat g^R \partial_t \hat f_1 - \partial_t \hat f_1 \widehat g^A )\} +  \\ \nonumber
     i\varepsilon [\hat\tau_3, (\widehat g^R \hat f_1 - \hat f_1 \widehat g^A )] 
      - \frac{1}{2} [ \widehat \Lambda,   \partial_\varepsilon \hat g^R \partial_t \hat f_1 +
          \partial_t\hat f_1  \partial_\varepsilon \hat g^A ] +
           \\ \nonumber
      i [ \widehat \Lambda, (\widehat g^R \hat f_1  -  \hat f_1 \widehat g^A )  ] +
     \frac{1}{2} \partial_\varepsilon f_0 \{ \partial_t \widehat \Lambda, \widehat g^{R} - \widehat g^{A} \} =St\{f_1\}.
     \end{align}
 %%%%%%%%%%%%%%%%%%%%%%%%%%%%%%%%%%%%%%%% 
 Here it is quite important to take into account modifications of spectral functions due to the Zeeman energy shift. 
 This modification has not been taken into account in previous work on the quasiclassical theory of spin dynamics in 
 superfluid $^3$He \cite{Eckern1981}. As discussed in recent works, 
 Zeeman spin splitting leads to the qualitative 
 changes in the quasiparticle spin transport 
 properties even in spin-singlet superconductors\cite{Bobkova2015,Silaev2015,SilaevRMP2017}.  
  
 Since the Zeeman shift is quite small as compared to the gap amplitude, it is enough to use the 
 first-order expansions in terms of the effective magnetic field 
 \begin{align} \label{Eq:GFZeemanExpansion}
 &\widehat g^R = \widehat g^R_0 + (\gamma/2) (\bm{h}\cdot \bm {d}_0) \widehat g^R_1 \\
 &\widehat g_0^{R} = \hat\tau_3 G_0 - i\hat\tau_1(\bm{\hat \sigma}\cdot  \bm{d}_0 ) F_0  \\
 &\widehat g_1^{R} = \hat\tau_3 (\bm{\hat \sigma}\cdot \bm{d}_0)  \partial_\varepsilon G_0 - i\hat\tau_1 \partial_\varepsilon F_0 ,
 \end{align}
 where $G_0= \varepsilon/\sqrt{\varepsilon^2-\Delta^2}$ and $F_0=\Delta/\sqrt{\varepsilon^2-\Delta^2}$. The advanced function
 is given by the usual relation $\widehat g^A = - \hat\tau_3 \widehat g^{R\dagger} \hat\tau_3$.
 Then after some algebra we transform the kinetic equation to the following form 
 \begin{align} \label{Eq:Kinetic}
  G_0\partial_t {\bm f}_T - \Delta \partial_\varepsilon F_0 \partial_t {\bm f}^\perp_T + \gamma G_0 ({\bm h}\times{\bm f}_T)
 -\gamma \Delta\partial_\varepsilon F_0 ({\bm d}_0\cdot{\bm h})({\bm d}_0\times{\bm f}_T )  = &
 \\ \nonumber
  \frac{\gamma}{2} G_0 \partial_\varepsilon f_0 \bm{\dot{h}} + St\{{\bm f}_T\} &   ,
 \end{align}
 where  we separate the transverse component of the distribution functions 
 with respect to ${\bm d}_0$ such that ${\bm f}_T^\perp={\bm f}_T - {\bm d}_0 ({\bm d}_0\cdot{\bm f}_T) $.  
 For the reasons discussed above, this quasiclassical equation (\ref{Eq:Kinetic}) 
 is different from that obtained in Ref. (\cite{Eckern1981}). 
 However it coincides with the one derived without using the quasiclassical approximation.
 To demonstrate that let us introduce the distribution function 
 \begin{equation} \label{Eq:SpinDistrFun}
 \bm{\nu}= {\bm d}_0 f_T^\parallel + G_0\bm{f}_T^\perp  
 \end{equation}
  Then, with the help of (\ref{Eq:Kinetic}) one obtains the Boltzmann-like kinetic equation
 \begin{equation} \label{Eq:Combescot}
 \partial_t {\bm \nu} -  2 \delta \bm{E} \times \bm{\nu} =
 -\partial_\varepsilon f_0 \partial_t( \delta \bm{E}) +  St\{{\bm \nu}_T\}   
 \end{equation}
 where the effective shift of energy levels under the action of the
  Zeeman field is given by
 \begin{equation}
 \delta \bm{E} = -\frac{\gamma}{2} \left( h_\parallel \bm {d}_0 +  \frac{{\bm h}_\perp}{G_0}\right) .
 \end{equation}
  Eq.(\ref{Eq:Combescot}) is identical to previous results obtained without quasiclassical approximation
 \cite{Combescot1974a,Combescot1974,Combescot1975,Combescot1976,Woelfle1976}.
  %%%%%%%%%%%%%%%%%%%%%%%%%%%%%%%%%%%%%%%%%%%%%%%%%%%%%%%%%%%%%%%%%%%%

 %%%%%%%%%%%%%%%%%%%%%%%%%%%%%%%%%%%%%%%%%%%%%%%%%%%%%%%%%%%%%%%%%%%%%%%%%%%%%%       
 \section{Spin relaxation due to surface Majorana states in superfluid $^3$He-B}
  The main simplification used in the derivation of the bulk kinetic equation is based on the 
  truncation of the gradient expansion to the first term in Eq.(\ref{Eq:gKgradientExp}). 
  This approximation is not valid to describe resonant transitions between 
  the energy levels corresponding to the surface bound states. These transitions happen at the 
  fixed momentum projection to the surface plane, so that the spectrum consists of the discrete 
      energy levels $\varepsilon_n$.  
  In this case the spectral functions have isolated poles 
  $\widehat{g}^{R,A} (\varepsilon)\propto \delta (\varepsilon - \varepsilon_n)$, so that e.g. the last term in the 
  expansion (\ref{Eq:gKgradientExp}) is much larger than  the first one. 
  However, the gradient expansion is still applicable in some cases with the discrete spectrum. For example, it 
  can be used for the description of localized fermionic states in the vortex cores
  \cite{Kopnin1991,Kopnin2002,KopninBook}. 
  The interlevel distance corresponding to the localized vortex core states is so small that the
   quasiclassical approximation yields continuous spectral branches.
    In case of the surface Andreev-Majorana states \cite{Volovik2010,SilaevVolovik2014} the situation is different since their spectrum 
  at fixed momentum projection is discrete even within the quasiclassics.
  Thus the gradient expansion is not applicable and it is not possible to derive the
  Boltzmann-like kinetic equation. 
    Below we will treat this problem by  finding the
     Keldysh function, which is a solution of the full Eilenberger equation (\ref{Eq:EilenbergerOrder1}),
  using the exact form of the spectrum and wave funtions near the surface. 
 %We consider spin dynamics near the surface of fully gapped superfluid $^3$He
 %with the normal vector along $\bm z$ axis. 
   
 %%%%%%%%%%%%%%%%%%%%%%%%%%%%%%%%%%%  
 \subsection{Andreev-Majorana bound states on the surface of $^3$He-B}
 In this section we derive the spectrum of localized states at the surface of $^3$He-B
 taking into account the Fermi-liquid corrections. 
 We assume that the surface normal is oriented along $\bm z$, the equilibrium order parameter 
 is defined by the rotation matrix $\hat R_0$ and 
 $\bm d_0 = ( q_x \Delta_\perp, q_y\Delta_\perp, q_z \Delta_\parallel)/\Delta $ 
 where the components depend on the distance to the surface 
 $\Delta_{\perp,\parallel} = \Delta_{\perp,\parallel}(z)$. 
  
 Then upon specular reflection from the surface the quasiparticle 
 momentum projection $q_z$
 and therefore the $z$-component of the order parameter 
 vector $\bm d_0$ change the signs which leads to the formation of
 surface bound states. Their spectrum is determined by the Andreev equation for the 
 quasiparticle wave 
 function $\hat\psi=\hat\psi (z)$, which is a four-component vector in spin-Nambu space
 \begin{equation}\label{Eq:Andreev}
 \hat H (z,\partial_z) \hat \psi =\varepsilon \hat\psi .
 \end{equation}  
 To find the discrete  spectrum of (\ref{Eq:Andreev}) we employ the 
 usual procedure of splitting the Hamiltonian in two parts 
  $\hat H=\hat H_0+\hat H_1$, so that
 \begin{align}\label{Eq:H0}
 &\hat H_0=-i\hat\tau_3 v_F q_z\partial_z +\hat\tau_2 \Delta_\perp \bm\hat\sigma_z q_z \\
 \label{Eq:H1}
 &\hat H_1=\Delta_\parallel\hat\tau_2 ( \bm{\hat \sigma}\cdot \bm{{q}}_\perp) - 
 \frac{\gamma}{2} (\bm{\hat \sigma} \cdot \hat R_0{\bm H}^{(0)}_{eff} ) ,
 \end{align}
 where $\bm{{q}}_\perp = (q_x,q_y,0)$ and $\hat R_0$ is the equilibrium order parameter rotation matrix. 
 The Hamiltonian (\ref{Eq:H0}) has zero-energy eigenvalues corresponding to the degenerate surface bound states.
 The correction from perturbation $\hat H_1$ results in the spectrum 
 \begin{equation}\label{Eq:Energy}
 \varepsilon_{1,2}=\pm \sqrt{C^2p^2_\perp + E_g^2/4},
 \end{equation}
  where $p_\perp = p_F q_\perp$. 
  The corresponding wave functions were found in 
  Refs. \cite{Volovik2010},\cite{Silaev2011a}.  
 The velocity and mass of the spectrum (\ref{Eq:Energy}) are given by
 \begin{align} \label{Eq:Velocity}
 & C=\frac{1}{p_F L_\xi }\int_0^{\infty} \Delta_\parallel  \exp\left[-2K(z)\right] dz 
 \\  \label{Eq:Gap}
 & E_g=\frac{\gamma}{L_\xi}\int_0^{\infty} (\bm n_s\cdot \bm{H}^{(0)}_{eff}) 
 \exp\left[-2K(z)\right] dz
 \end{align}
 where $\bm{n}_s = \hat R_0 \bm{\hat z}$ is the spin quantization axis, 
  $K(z) = \frac{1}{\hbar v_F}\int_0^z \Delta_\perp (z) dz $,
 $L_\xi = 4\int_0^\infty e^{-2K(z)}dz  $ is a normalization length which 
 is of the order of the coherence length
 $\xi = \hbar v_F /\Delta$.
 
 The minigap $E_g$ is induced by the external magnetic field.
 Here we point out that Fermi-liquid corrections lead to the strong re-normalization of 
 $E_g$ as compared to its 'bare' value given by $E^{(0)}_g = \hbar \omega_L$, where $\omega_L=\gamma H_{ext}^{(0)}$
 is the Larmor frequency.
 The effective field in (\ref{Eq:Gap}) is given by (\ref{Eq:Heff}) which can be written  
 in terms of the local spin susceptibility  $ \chi = \chi (z)$ so that
 $\bm{H}^{(0)}_{eff} (z)= {\bm H}_{ext} (1 -  (Z_0/4) \chi/\chi_{n0}) $.
 To calculate an exact value of $E_g$ one has to determine $\chi(z)$ self-consistently
 \cite{Mizushima2012}. For the estimation we can assume that $\chi \approx \chi_n$, 
 where $\chi_n= \chi_{n0}/(1+Z_0/4)$
 is the renormalized bulk normal-state susceptibility. Then for $Z_0\approx -3$ the spectral minigap is given by 
 $E_g \approx 4\hbar \omega_L $. As we will see below at smaller frequencies $\omega<E_g/\hbar$
 surface bound states give no absorption signal. In particular this situation is realized for the experimentally interesting 
 domain of frequencies in the close vicinity of the main NMR peak located at the Larmor frequency $\omega\approx \omega_L < E_g/\hbar$.  
  %As will be shown below in this regime one can neglect the dispersion of the SMS magnetic response. 
  
  The localized states of the Andreev equation (\ref{Eq:Andreev})
  with discrete spectrum $\varepsilon_n$ provide singular contributions to the Green's functions:
%  %
%  \begin{equation} \label{Eq:GFandreev}
%  \widehat G^{R,A} (z_1,z_2, \bm{q}_\perp, \varepsilon) = 
%  \sum_n \frac{|\hat\psi_n (z_1)\rangle \langle \hat\psi_n (z_2)| }{ \varepsilon -\varepsilon_n \pm i 0} .
%  \end{equation}
%  %
  %
  \begin{equation} \label{Eq:GFandreev}
  \widehat G^{R,A} (z_1,z_2, \bm{q}_\perp, \varepsilon) = 
  \sum_n \frac{|\hat\psi_n (z_1)\rangle \langle \hat\psi_n (z_2)| }{ \varepsilon_n - \varepsilon \mp i 0} .
  \end{equation}
  This expression will be used below to calculate the singular part of 
  quasiclassical propagators determined by the contribution
  of surface Andreev-Majorana states. 
    
  %%%%%%%%%%%%%%%%%%%%%%%%%%%%%%%%%%%%%%%%%%%%%%%%%%%%%%%%%%%%%%%%%%%%%%  
 \subsection{Quasiclassical propagators}
 To proceed we need to calculate spectral functions $g^{R,A}$ near
 the surface of $^3$He-B. We will use a general relation \cite{KopninBook} for the singular part of quasiclassical
  propagators in terms of the Green's functions (\ref{Eq:GFandreev}) of Andreev equations
 \begin{equation}\label{Eq:RelationGg}
 (\widehat g^{R}-\widehat g^{A})(z) = -2i v_{z} \hat\tau_3 \left (\widehat G^R-\widehat G^A\right)
 \end{equation}
 where $v_{z}=(\bm v_F\cdot  \bm {\hat z})$ and
 $\widehat G^{R,A} = \widehat G^{R,A} (z_1=z_2=z, \bm{q}_\perp, \varepsilon)$ are the Green's
 functions of Andreev equations (\ref{Eq:Andreev}) which depend on the coordinate $z$
 perpendicular to the surface, momentum projection on $xy$ plane
 $\bm {\hat p}_\perp$ and energy $\varepsilon$. 
  We are interested in the contribution of
 bound states and therefore can use the expression
 $ (\widehat G^{R}- \widehat G^{A})(z_1=z_2)= 2\pi i\; \sum_n \delta(\varepsilon - \varepsilon_n) |\hat\psi_n \rangle \langle\hat\psi_n| $,
 so that 
 \begin{equation}\label{Eq:gRApoles}
 \widehat g^{R}- \widehat g^{A} = 4\pi v_z \sum_n \delta(\varepsilon - \varepsilon_n) |\hat\tau_3\hat\psi_n \rangle \langle\hat\psi_n| .
 \end{equation} 
The equilibrium Keldysh function is given by 
$\widehat g^K (\varepsilon)= f_0(\varepsilon)( \widehat g^{R}-\widehat g^{A})$,
 where $f_0(\varepsilon)=\tanh(\varepsilon/2T)$ is the equilibrium distribution function. Hence in the time domain
  we get  $\widehat g^{K} (t,t^\prime) = \int \widehat g^{K} (\varepsilon) \exp(i\varepsilon (t^\prime- t)) d\varepsilon $
  so that
  \begin{equation}\label{Eq:gKpoles}
 \widehat g^{K}_{eq} (t,t^\prime)= 4\pi v_z \sum_n e^{i\varepsilon_n (t^\prime- t)} 
 f_0(\varepsilon_n) |\hat\tau_3\hat\psi_n \rangle \langle\hat\psi_n| .
 \end{equation}

 %%%%%%%%%%%%%%%%%%%%%%%%%%%%%%%%%%%%%%%%%%%%%%%%%%%%%%%%%%%%%%%%%%% 
 \subsection{Non-equilibrium spin surface states } 
 \label{Sec:NonEquilibriumSpinSurface}
 Having in hand the equilibrium Keldysh function (\ref{Eq:gKpoles}) we can proceed to study non-equilibrium 
 spin polarization of surface bound states given by the general kinetic equation (\ref{Eq:EilenbergerOrder1}). 
 We search for the first order correction 
 to the Keldysh function $\widehat g^K_1 = \widehat g^K_1 (t,t^\prime)$ substituting the equilibrium function in the form (\ref{Eq:gKpoles})
 to the r.h.s. of Eq.(\ref{Eq:EilenbergerOrder1}). 
 Assuming that the driving field is $\bm { h}(t) = {\bm h}_{\omega}e^{i\omega t} $  
  and neglecting the collision integral we can find the analytical solution of 
  Eq.(\ref{Eq:EilenbergerOrder1}) in the following form
 \begin{align} \label{Eq:Solution}
 & \widehat g^K_1 (t,t^\prime) =  2\pi\gamma v_z \sum_{n\neq m} 
 \frac{\langle \psi_n| \bm h_\omega\cdot \bm{\hat \sigma} | \psi_m \rangle}{\omega + \varepsilon_n -\varepsilon_m} \times
   \\ \nonumber
 & \left( 
 f_0(\varepsilon_m) e^{i\varepsilon_m t^\prime + i(\omega - \varepsilon_m) t}
  - 
 f_0(\varepsilon_n) e^{i(\varepsilon_n +\omega) t^\prime - i\varepsilon_n t}
 \right)   | \tau_3\psi_n \rangle\langle \psi_m | ,
 \end{align}
 where $\bm{h}_\omega$ denotes the gauge-invariant 
  effective field obtained from Eq.(\ref{Eq:TimeDependentField})
  \begin{equation}\label{Eq:GaugeInvariantField1}
  \bm{h}_\omega = 
   \bm{\tilde H}_{eff} + {\bm \theta}\times{\bm H}^{(0)}_{eff} - i\omega \bm {\theta}/\gamma.
  \end{equation} 
  Calculating the Fourier component $\widehat g^K_1 (\omega) =\int e^{-\omega t} \widehat g^K_1 (t,t^\prime) dt$, 
  using the definition (\ref{Eq:Spin}) and the matrix element 
  $ \langle \psi_1| \bm{\hat \sigma} | \psi_2 \rangle = (Cp/\varepsilon_1) \bm{n}_s $  
  we obtain the frequency-dependent spin polarization of Andreev-Majorana bound states per unit surface area 
  \begin{equation} \label{Eq:SpinRotating}
  \bm{S}_{bs}(\omega) = \frac{\chi_{bs}(\omega)}{\gamma}(\bm{n}_s\cdot \bm{\bar h}_\omega ) \bm{n}_s ,
  \end{equation}
  where $\bm{n}_s$ is the spin quantization axis of surface states and 
  \begin{equation}
  \bm{\bar h}_\omega = L_\xi^{-1}\int_0^{\infty} \bm{h}_\omega (z)\exp\left[-2K(z)\right] dz 
  \end{equation}    
  is the driving field (\ref{Eq:GaugeInvariantField1}) 
  averaged over the surface bound state localization scale. In the absence of Fermi-liquid corrections 
  $\bm{\bar h}_\omega = \bm{h}_\omega$.
  The longitudinal susceptibility $\chi_{bs}(\omega)$ coincides with the expression found in 
  Refs. \cite{Silaev2011a}, \cite{Taylor2015} 
 \begin{equation}\label{Eq:ZZ1}
 \chi_{bs}=\frac{\gamma^2C^2}{8\pi\hbar^2} \int_0^{p_F} p^3d p
 \frac{(\varepsilon_2-\varepsilon_1)\left[f_0(\varepsilon_1)-f_0(\varepsilon_2)\right]}
 {\varepsilon_1^2[ (\hbar \omega )^2-(\varepsilon_2-\varepsilon_1)^2]} .
  \end{equation}
  The dissipation rate is determined by the imaginary part of
  $\chi_{bs}$, which is non-zero at the frequencies larger than the minigap  
  $\omega>E_g/\hbar$
 \begin{equation}\label{Eq:ImZZ}
  Im \chi_{bs}= - \left(\frac{\gamma}{4\hbar C}\right)^2\frac{\hbar\omega}{2}f_0\left(\frac{\hbar\omega}{4T}\right)
  \left(1-\frac{E_g^2}{\hbar^2\omega^2}\right).
 \end{equation}
  In contrast to the previous considerations \cite{Silaev2011a,Taylor2015} which neglected Fermi-liquid corrections to the spectrum, 
  the result (\ref{Eq:ImZZ}) demonstrates that the absorption threshold  is given by $E_g/\hbar$
  which is significantly larger than the Larmor frequency as discussed above.   
  
 %%%%%%%%%%%%%%%%%%%%%%%%%%%%%%%%%%%%%%%%%%%%%%%%%%%%%%%%%%%%%
 \subsection{Gauge-invariant theory of surface relaxation}
 \label{SubSec:GaugeInvariantSurface}
 Let us consider the influence of the transitions in the spectrum of surface states on the dissipation rate of
 spin waves in the fully gapped superfluid $^3$He-B.  
 In general the dynamics of the total spin in the laboratory frame is governed by Eq.(\ref{Eq:SpinConservation}). 
 Let us consider a monochromatic signal $\bm{ S} \propto e^{i\omega t}$ to obtain from (\ref{Eq:SpinConservation})
 \begin{equation} \label{Eq:SpinConservation1}
  \nabla_k \bm J_k + i\omega\bm{ S} = \bm{S}\times\bm{\omega}_L + \bm{R}_D,
 \end{equation}
 where $\bm{\omega}_L = \gamma \bm{H}_{ext}$ and the dipole interaction $\bm{R}_D$ is a function of the rotation 
 vector $\bm{\theta} $  which parametrizes deviations of the order parameter from equilibrium 
 determined by the rotation matrix (\ref{Eq:RotationMatrix}).
  In the bulk B-phase the dipole torque is given by \cite{WolfleBook}
 \begin{equation} \label{Eq:DipoleTorque1}
 \bm{R}_D = \frac{\chi_B\Omega^2_B}{\gamma} \bm{n}(\bm{n}\cdot\bm{\theta}),
 \end{equation}
 where $\Omega_B$ is the B phase longitudinal resonance frequency \cite{Leggett1975a,WolfleBook} and
  $\chi_{B}$ is the B phase bulk susceptibility.
 The total spin density $\bm S = \bm S (\omega, z)$ is given by the superposition of
 the bulk contribution and that of the Andreev-Majorana bound states, localized at the distance of the 
 coherence length $\xi$ near the surface. Since all length-scales which determine spin dynamics in the bulk 
 are much larger than $\xi$, we can write $\bm S$ in the rotating frame as follows  
 \begin{equation} \label{Eq:SpinBulkSurface}
 \bm{S}(\omega,z) = \frac{\chi_{B0}}{\gamma} \bm h_\omega + \bm{S}_{bs} (\omega) \delta (z) ,
 \end{equation}
  where $\chi_{B0}$ is the bulk susceptibility without Fermi-liquid corrections, while the
  amplitude of the second term is given by Eq.(\ref{Eq:SpinRotating}).
  
  The main difficulty for the further analytical calculations is the influence of the 
  Fermi-liquid corrections on the spin of localized states in Eq.(\ref{Eq:SpinRotating}).
  To obtain qualitative results we neglect these corrections in the surface term and keep 
  them for the bulk contribution.
  In this case the gauge-invariant driving field is given by its 'bare' form
  $\bm{h}^{(b)}_\omega \approx \bm{\tilde H}_{ext} + {\bm \theta}\times{\bm H}^{(0)}_{ext} - i\omega \bm {\theta}/\gamma$.
   Hence we will use the approximate expression for the spin signal 
  \begin{equation} \label{Eq:SpinBulkSurface2}
 \bm{S}(\omega,z) = \frac{\chi_{B}}{\gamma} \bm{h}^{(b)}_\omega + \bm{S}_{bs} (\omega) \delta (z) ,
 \end{equation}
  where $\chi_{B}$ is the total B-phase bulk susceptibility and
  the second term is given by Eq.(\ref{Eq:SpinRotating}) with $\bm{h}^{(b)}_\omega $ instead of $\bm{ h}_\omega $.

 \subsubsection{Longitudinal resonance in thin film of $^3$He-B} 
 \label{SubSec:LongitudinalResonance}
 First we consider the influence of Andreev-Majorana states on the decay of 
 longitudinal modes \cite{WolfleBook} when ${\theta} = \theta_z {\bm z}$.
 For simplicity we assume that superfluid is homogeneous without any underlying texture so that 
 the rotation axis of the matrix $\hat R$ is directed along the surface normal
 ${\bm n}\parallel {\bm z}$. 
 In this case the spin density can be written as $\bm{S}(\omega,z) = \tilde\chi (z) \bm{h}^{(b)}_\omega$, where 
 \begin{equation} \label{Eq:LocalChi}
 \tilde\chi (\omega,z) = \chi_{B} + \chi_{bs}(\omega) \delta (z). 
 \end{equation} 
 Note that $\tilde\chi$ is not the spin susceptibility, since the driving field $\bm{h}^{(b)}_\omega$
 contains the dynamical variable $\bm\theta$ in addition to the external field. 
 In order to find the 
  true susceptibility $\hat\chi$ which relates the total spin density and external magnetic field
  $\bm S= \hat\chi \bm H_{ext}$ the angle $\bm\theta$ has to be determined from the general 
  equation for the total spin dynamics (\ref{Eq:SpinConservation1}).
  Then writing for the spin current $\nabla_z \bm J_z = c_z^2\nabla_z^2 \theta_z$, 
  where $c_z$ is 
 the spin wave velocity, and combining Eqs.(\ref{Eq:SpinConservation1},\ref{Eq:DipoleTorque1},\ref{Eq:SpinBulkSurface}) we get 
 \begin{equation}\label{Eq:thetaZ}
 \left( c_z^2\nabla_z^2 + \Omega_B^2 - 
 \omega^2 \frac{\tilde\chi}{\chi_B} \right )\theta_z 
  = 
 i\omega \gamma \tilde H_{ext}\frac{\tilde\chi}{\chi_B} .
 \end{equation}   
  The mode with the smallest frequency $\omega_0 \approx \Omega_B$ is given by the 
  space-homogeneous state $\theta_z (z)=const$.
  Its life time is given by the imaginary part of the frequency which can be found 
  by averaging over the coordinate $z$ the homogeneous Eq. (\ref{Eq:thetaZ}) 
  with $\tilde H_{ext} =0$ 
 \begin{equation} \label{Eq:RelaxationLongitudinal}
 \tau^{-1} = - \frac{ \Omega_B{\rm Im}\chi_{bs}}{2L}, 
 \end{equation}
  where $L$ is the film thickness. 
  In case if there are surface bound states on both surfaces of the film
  the relaxation rate (\ref{Eq:RelaxationLongitudinal}) is doubled. 
 
 The longitudinal magnetic susceptibility $\chi_{zz}$ which relates 
 the total spin density and external magnetic field
  $ S = \chi_{zz} H_{ext}$ can be found using Eqs. (\ref{Eq:RelaxationLongitudinal}) and 
 (\ref{Eq:SpinConservation1}): 
 \begin{equation}\label{Eq:SusceptibilityFin}
 \chi_{zz} = \bar\chi \frac{\chi_B \Omega_B^2 }{ \chi_B \Omega_B^2 - \omega^2\bar\chi  },
 \end{equation}
 where $ \bar\chi$ is the function (\ref{Eq:LocalChi}) averaged over the film thickness. In the absence of 
 Fermi-liquid corrections this result coincides with the one obtained before \cite{Taylor2015} from the effective action approach. 
 The most significant difference is however in the behaviour of the 
 bound states spin susceptibility ${\rm Im} \chi_{bs}$ given by (\ref{Eq:ImZZ}) which becomes finite at the frequencies 
 larger than $E_g\approx 4 \hbar \omega_L$ rather than at the bare Larmor frequency. 
  
 %%%%%%%%%%%%%%%%%%%%%%%%%%%%%%%%%%%%%%%%%%%%%%%%%%%%%%%%%%%%%%%%%%
 \subsubsection{Transverse resonance in a texture}
 \label{SubSec:TranserceResonance}
 The basic measurement tools in superfluid $^3$He experiments are frequency shifts and dissipation
 rates of the transverse magnetic precession modes in the presence of 
 a large static magnetic field component. Recently, the technique based on the relaxation of 
 a magnon condensate has been developed \cite{Fisher2012,Autti2012,Heikkinen2014,Heikkinen2014a,EltsovArxiv2013,Zavjalov2016}.
 In principle, it can be used for the identification of surface Andreev-Majorana 
 states although 
 this approach has several difficulties discussed below.
  
  Let us assume that the constant magnetic field is directed along the surface normal 
  $\bm H_{ext}^{(0)} \parallel \bm z$
  and the non-equilibrium spin state is driven by the time-dependent perpendicular component 
  $\bm {\tilde H}_{ext} \perp \bm H_{ext}^{(0)}$. If the rotation axis is parallel to the constant field
  $\bm n\parallel \bm z$  and the spin quantization axis is  $\bm n_s\parallel \bm z$  
  then according to Eq. (\ref{Eq:SpinRotating}) the oscillating transverse field component cannot 
  induce transitions of the surface bound states. Therefore the presence of 
  Andreev-Majorana states shows up in the transverse resonance only 
  if $\bm n$ is deflected with a finite angle $\beta_n$ from the $\bm z$-axis
  so that the effective driving field $\bm h$ has a component parallel to the 
  spin quantization axis $\bm n_s$.
  Physically, the texture of $\bm n$ can appear via the interplay between gradient and 
  dipole energies. However in typical experimental setups \cite{EltsovArxiv2013} the 
  angle $\beta_n$ is rather small near the surface which leads to a strong suppression of  
  the spin response from the surface states. 
  
  To quantify the effect of surface bound states on the transverse resonance we
  can use the general theory described in Sec. (\ref{SubSec:GaugeInvariantSurface}).
  Here we need to take into account that the expression for the non-equilibrium spin density
  Eq. (\ref{Eq:SpinBulkSurface2}) is obtained in the rotating frame, while
  Eq. (\ref{Eq:SpinConservation1}) is written in the laboratory frame. 
  Hence the non-equilibrium spin density has to be rotated back to the laboratory frame    
  $\hat R^{-1}_{ac} \bm{S} \approx \bm{S}  + \bm{S}\times {\bm \theta}$ so that Eq.(\ref{Eq:SpinBulkSurface2}) changes as follows
  \begin{equation} \label{Eq:SpinBulkSurface1}
  \bm{S}(\omega,z) = \frac{\chi_{B}}{\gamma} (\bm{\tilde H}_{ext}-i\omega\bm\theta/\gamma) + \bm{S}_{bs} (\omega) \delta (z) ,
  \end{equation}  
  where the surface contribution  $\bm{S}_{bs} $ remains unchanged given by the 
  Eq. (\ref{Eq:SpinBulkSurface}) 
  with the driving field $\bm{h}^{(b)}_\omega$. 
  To describe the transverse resonance which occurs at frequencies close to the Larmor frequency  $\omega\approx \omega_L$ 
  we project the above general expressions (\ref{Eq:SpinConservation1},\ref{Eq:DipoleTorque1},\ref{Eq:SpinBulkSurface1}) 
  on the spin state corresponding to the optical magnons \cite{Autti2012,Zavjalov2015,Zavjalov2016}
  $ \bm{\theta} = \Psi \bm s_{opt} $, where $\Psi$ is the complex amplitude which can be 
  considered as a wave function and the polarization vector is $\bm s_{opt} = (1, i, 0)/\sqrt{2}$. 
    
  To find the relaxation time of transverse optical magnons due to the excitation of surface 
  bound states we collect Eqs. (\ref{Eq:SpinConservation1},\ref{Eq:DipoleTorque1},\ref{Eq:SpinBulkSurface1}),
  put $\bm{\tilde H}_{ext} = 0$ to obtain the following equation for the wave function
  \begin{equation} \label{Eq:Shr}
  \omega\Psi=  (\hat L + U)\Psi +  \frac{\chi_{bs} (\omega-\omega_L)^2}{2\chi_B\omega_L} \sin^2\beta_s \Psi \delta (z).
  \end{equation}
 Here $\sin\beta_s = \sqrt{ 1 - (\bm n_s \bm z)^2 }$  and $\hat L$ denotes the gradient terms coming from the spin current divergence,  
 $U$ is the effective potential energy
 \begin{align} \label{Eq:L}
 & \hat L = - \frac{\hbar\nabla_\perp^2}{2m_\perp} -  \frac{\hbar\nabla_z^2}{2m_z} 
 \\ \label{Eq:U}
 & U=\frac{\Omega_B^2}{2\omega_L} \sin^2\beta_n  + \omega_L ,
 \end{align} 
 where $\sin\beta_n = \sqrt{ 1 - (\bm n \bm z)^2 }$ . 
 Effective masses $m_z = \hbar \omega_L/2c_z^2$ and $m_\perp = \hbar \omega_L/2c_\perp^2$
 are related to the longitudinal $c_z$ and transverse $c_\perp$ spin wave velocities.  
  Apart from the surface-related last term on the r.h.s., Eq. (\ref{Eq:Shr}) coincides with the 
 equations considered before to describe spin waves on the optical branch modified by the 
 $\bm n$-vector texture \cite{Autti2012,Zavjalov2015}.
 The surface term in (\ref{Eq:Shr}) yields a finite lifetime of the spin waves at the frequencies where 
 ${\rm Im} \chi_{bs}\neq 0$.   
   To quantify the relaxation effect we calculate the inverse lifetime of 
   the magnon with frequency $\omega_n$ with the spatial distribution of 
   spin described by the wave function $\Psi_n$    
   \begin{equation} \label{Eq:Relaxation}
  \tau^{-1} = \frac{{\rm Im}\chi_{bs} }{\chi_B}\frac{(\omega_n - \omega_L)^2}{\omega_L} 
  \frac{\int_S d^2r \sin^2\beta_s  |\Psi_n|^2}{\int_V d^3r  |\Psi_n|^2}
  \end{equation}      
  where $S$ and $V$ denote the surface and the volume of $^3$He-B. 
  Bearing in mind that NMR occurs at frequencies are rather close to the Larmor frequency 
  $\omega_n \approx \omega_L$, one can see that minigap renormalization by Fermi-liquid corrections (\ref{Eq:Gap})
  makes it impossible to excite Andreev-Majorana states by the transverse 
  magnetic resonance.  
  Indeed in this case the absorption threshold is shifted to 
  the frequencies $E_g/\hbar \sim 4 \omega_L$ much larger than the driving frequency of the magnetic precession
  $\omega_n\approx \omega_L$. 

  However, 
  Majorana states can be excited by the transverse spin waves which have the frequencies higher that $E_g$
  and correspondingly the wavelengths of the order of $v_F/E_g$. 
    Provided that $E_g\ll \Delta$ the required wavelengths are much larger than the coherence length $\xi$, 
   which determines the localization scale of the surface bound states. 
   Therefore, to describe the interactions of Majorana states with these spin waves 
   one can consider the magnetization precession  which is locally homogeneous in space and 
    use the same equations for the fermionic spin response as considered above. 

 \section{Conclusions} 
 To conclude, we have presented the quasiclassical theory of spin dynamics in superfluid $^3$He.
 Starting from the most general quasiclassical Kedlysh-Eilenberger equation we derived the kinetic equation for the
 spin-dependent distribution function in the bulk phase. The result coincides with that obtained 
 some time ago without quasiclassical approximation. 
 Applying this technique we have obtained the frequency-dependent lifetimes of longitudinal and transverse 
 spin waves due to their interaction with  Andreev-Majorana states localized at the surface of $^3$He-B phase. 
 With the help of the quasiclassical approach the crucial role of Fermi-liquid corrections in the magnetic response 
 is demonstrated. 
   An important qualitative conclusion is that such relaxation mechanism can be effective only for the 
   longitudinal spin resonance. At the same time, the spatially-homogeneous transverse NMR mode does not 
   excite Andreev-Majorana surface because of two reasons.  First, due to the Fermi-liquid corrections the minigap 
   in the  surface state spectrum is much larger than NMR frequency which is close to the Larmor frequency.
    Second, the matrix element of transitions in the Andreev-Majorana spectrum 
   is proportional to the deflection angle $\beta_s$ of the spin quantization axis from the constant magnetic field component. 
   Usually the texture of the vector $\bm n$  is flat near the surface so that 
   the  deflection is rather small, $|\beta_{s}|\ll 1$.
 The finite-momentum transverse spin waves with frequencies larger than $E_g$
can excite Majorana states. The interaction of such spin waves with the surface states can be 
      described within the same theoretical framework as considered in this paper.
           
  \section{Acknowledgements}
  We thank V. El'tsov and G. Volovik for stimulating discussions. 
  This work was supported by the Academy of Finland.
        
 % \bibliographystyle{spphys}       % APS-like style for physics
 % \bibliography{LiteratureJLTP}   

\end{document}